\documentstyle[11pt]{article}
\begin{document}
\twocolumn

{\noindent\bf  Disk contamination in a sample of proper motion selected halo stars}\\
{A comment on {\it Direct Detection of Galactic Halo Dark Matter}, B.A. Oppenheimer et al. 2001, Science}
\vspace{0.1in}

{\noindent David S. Graff --- {\tt graff@umich.edu}}\\
{\it University of Michigan Astronomy\\
 Ann Arbor, MI 48109}
\vspace{0.1in}

{\noindent submitted as a Technical Comment to {\it Science}}\\

\vspace{0.1in}
\noindent Sir:

I am writing in comment to the recent survey by Oppenheimer et al
\cite{opp}.  This survey finds 34 high velocity white dwarfs, and
concludes that these white dwarfs form a component of dark matter,
representing 3\% of the local mass of the halo.

This survey is similar to the Luyten survey \cite{luyten}, covering
10\% of the sky whereas Luyten covers 2/3 of the sky, but this survey
goes somewhat deeper.  It covers about 2/3 the volume of the Luyten
survey, ignoring the proper motion limits.  The authors are sensitive
to a similar range of proper motions to Luyten, though the range of
velocities is not the same since the Oppenheimer et al. survey will
see objects to a larger distance, and hence a larger velocity.

Since the surveys are so similar, it is surprising that the results
are so different.  However, closer examination shows that the results
are not different.  A further analysis of the Luyten survey including
trigonometric parallax \cite{ldm} shows that roughly half Luyten's
white dwarfs have velocities greater than 94 km s$^{-1}$, as is true
for Oppenheimer et al's.  Whereas these objects were classified as
disk or thick disk dwarfs by \cite{ldm}, they are classified as Halo
objects by Oppenheimer et al.

In both surveys, these stars are the high velocity Gaussian tail of
the disk population.  The density of halo main sequence stars is about
1/600 the density of disk stars.  The authors claim that their cut at
94 km/sec will eliminate 95\% of the disk population, but this means
that the remaining 5\% of the disk dwarfs should still outnumber the
halo dwarfs by 30 to 1.

The problem is even worse since this is a proper motion limited
survey.  Such a survey obviously selects for the highest velocity
stars, thus the typical velocity of a proper motion limited survey will be
higher than in the underlying population.  A quick Monte Carlo
simulation shows that the mean velocity in a proper motion limited
survey is about twice that of the underlying population.  Thus, the 94
km/sec threshold is actually a 1 $\sigma$ cutoff, and we expect a fair fraction
of detected disk stars pass this threshold.  This is born out by
examining figure 3 of Oppenheimer et al. which shows that most of the
white dwarfs in the sample are rotating around the galaxy in the same
direction as disk stars (on the right half of the diagram.

The authors have noticed this weighting in their sample, but claim
that this is due to a selection effect: they are not sensitive to the
faster moving halo stars.  The upper limit to proper motions is 3
$^{\prime \prime}$/yr.  Only one star approaches this limit, F351-50 at 2.5
$^{\prime \prime}$/yr.  The second highest proper motion is 1.7$^{\prime \prime}$/yr and the
third highest is $1.1^{\prime \prime}$/yr.  

This paper does not discover a new dark matter population, but only
rediscovers the white dwarfs of the disk.

\end{document}